\documentclass[twocolumn,showpacs,prd,floatfix,axodraw]{revtex4}
\usepackage{mathrsfs}
\usepackage{graphicx,,booktabs,bm}
\usepackage{overpic}
\usepackage{color}
\usepackage{amssymb}

\usepackage{mathrsfs,bm,amsmath,amssymb}
\usepackage{longtable,lscape}
\usepackage{txfonts}
\usepackage{amssymb}
\usepackage{indentfirst}
\usepackage{graphicx,,booktabs}
\usepackage{multirow,ulem}

\usepackage{color}
\usepackage{amssymb}

\definecolor{cover}{rgb}{0.77,0.87,0.88}
\definecolor{blueone}{rgb}{0.1,0.1,.7}
\definecolor{citec}{rgb}{0.14,0.47,0.09}
\definecolor{two}{rgb}{0.0,0.5,0.}
\definecolor{three}{rgb}{.5,.1,0.15}
\usepackage[bookmarks=true,bookmarksopen=false,plainpages=false,breaklinks=true,
   bookmarksnumbered=true,hypertexnames=false,
   filecolor=blue,urlcolor=three,menucolor=three,
   linkcolor=three,citecolor=blueone, colorlinks,
   anchorcolor=blue,runcolor=pink,frenchlinks=red
   pdfstartview=FitH,pdftitle=title,%
   pdfauthor=author]{hyperref}

\begin{document}
\title{Searching for the $G(3900)$ via the $K^- p \to D_s^- \Lambda_c^+ G(3900)^0$ reaction}
\author{Qing Lu}
\affiliation{Key Laboratory of Computational Physics of Sichuan Province, School of Mathematics and Physics, Yibin University, Yibin 644000, China}
\author{Cai Cheng}
\affiliation{School of Physics and Electronic Engineering, Sichuan Normal University, Chengdu 610101, China}
\author{Yin Huang\footnote{corresponding author}} \email{huangy2019@swjtu.edu.cn}
\affiliation{School of Physical Science and Technology, Southwest Jiaotong University, Chengdu 610031,China}

\begin{abstract}
The nature of the $G(3900)$ structure, observed in $e^{+}e^{-}\to D\bar{D}$, remains unclear and may stem either from a genuine resonance or from charmonium interference and threshold effects.
We therefore propose searching for the $G(3900)$ signal in the reaction $K^- p \to D_s^- \Lambda_c^+ G(3900)^0$, where the interference effects present in $e^{+}e^{-}\to \bar{D}^{*}D$ are absent.
We employ an effective Lagrangian approach, where the reaction proceeds via a central production mechanism dominated by $t$-channel $D^{0}$ and $D^{*0}$ exchanges, based on the possible interpretation
of $G(3900)$ as a $P$-wave $\bar{D}^{*}D$ molecular state, whose coupling to the $\bar{D}^{*}D$ channel is fixed from our previous fit to the $e^{+}e^{-}\to \bar{D}^{*}D$ data.  The $\bar{K}N$ initial-state
interaction, mediated by Pomeron and Reggeon exchanges, is also included and leads to a significant enhancement of the production cross section.  If measured in future experiments, the predicted total
cross sections and angular distributions can provide a promising probe of the nature of the $G(3900)$, and in particular of its possible genuine resonance nature.
\end{abstract}
\date{\today}

%\pacs{{13.60.Le}{inner structures}   \and {13.85.Lg}{Schr\"{o}dinger equation} \and  {25.30.-c}{molecular state}  }}

\maketitle
\section{Introduction}\label{sec:intro}
In 2024, the BESIII Collaboration analyzed an enlarged $e^{+}e^{-}\to D\bar{D}$ data sample and reaffirmed the existence of the $G(3900)$ resonance~\cite{BESIII:2024ths}.
The measured resonance parameters were determined to be $M_{G(3900)} = 3872.5 \pm 14.2 \pm 3.0~\mathrm{MeV}$ and $\Gamma_{G(3900)} = 179.7 \pm 14.1 \pm 7.0~\mathrm{MeV}$.
These results are reviewed in detail by Wang et al~\cite{Wang:2025dur}. 
Evidence for this resonance had previously been reported independently by the BaBar Collaboration~\cite{BaBar:2006qlj} and the Belle Collaboration~\cite{Belle:2007qxm} in
2007 and 2008, respectively. Its existence, however, remained highly controversial at the time. Some studies argued that the observed enhancement did not correspond to a
genuine resonance~\cite{Eichten:1979ms,Zhang:2009gy}, but instead arose from threshold effects associated with the opening of the $D\bar{D}^{*}+D^{*}\bar{D}$ channels,
together with the nodal structure of the $\psi(3S)$ radial wave function~\cite{Eichten:1979ms}.  The analyses of the existing experimental data~\cite{BaBar:2006qlj,Belle:2007qxm}
in Ref.~\cite{Du:2016qcr}, however, demonstrated that a satisfactory description of the observed $e^{+}e^{-}\to D\bar{D}$ line shape requires the explicit inclusion of the
$G(3900)$ contribution, thereby lending support to its interpretation as a distinct hadronic structure.

The 2024 BESIII observation reignited intense debate over the existence and nature of the $G(3900)$ resonance.  As in earlier studies, two competing interpretations remain.
One interprets the $G(3900)$ as a $P$-wave hadronic molecular state~\cite{Lin:2024qcq}, a scenario that can successfully reproduce the experimental data~\cite{Ye:2025ywy}.
The analysis based on a more complete data set provides additional support for the interpretation of $G(3900)$ as a $D\bar{D}^{*}$ molecular state~\cite{Nakamura:2023obk}.
The alternative interpretation~\cite{Husken:2024hmi} is based on a coupled-channel analysis of the latest experimental data, similar to that of Ref.~\cite{Ye:2025ywy}. However,
using the $K$-matrix approach instead of the method adopted in Ref.~\cite{Ye:2025ywy}, it arrives at the opposite conclusion, finding no evidence that the $G(3900)$ corresponds
to a genuine resonance.  Using an interaction potential that incorporates six coupled channels, Ref.~\cite{Salnikov:2024wah} solved the radial Schr\"{o}dinger equation and
compared the resulting line shapes with the experimental data, likewise finding no support for the interpretation of the $G(3900)$ as a genuine resonance state.

Therefore, clarifying whether the $G(3900)$ is a genuine hadronic resonance has become an active topic of current research, with a variety of approaches being employed to address
this question.  A PACIAE+DCPC simulation of $e^+e^-$ collisions shows that the yield, rapidity, and transverse-momentum distributions of $G(3900)$ differ significantly from those
of $X(3872)$ and $Z_c(3900)^0$, suggesting that these observables may help determine whether $G(3900)$ is a genuine resonance~\cite{Cao:2025okk}.  We further propose two experimental
signatures of a genuine $P$-wave molecular $G(3900)$. The first is a triangle-singularity-induced enhancement associated with its production mechanism~\cite{Huang:2025rvj}. The second,
and more direct, is a distinct Jacobian peak in the transverse-momentum distribution of the final-state $D$ mesons in the $e^{+}e^{-}\to D^{-}D^{+}$ reaction~\cite{Huang:2025xvv}.
Observation of such a peak would provide strong evidence for the existence of the $G(3900)$ state.

Furthermore, in Ref.~\cite{Lu:2025zae}, the same complex scaling method as that used in Ref.~\cite{Lin:2024qcq} is employed to solve the non-relativistic Schr\"odinger equation for
both bound states and resonances, with the difference that only the $t$-channel interaction is taken into account. The resulting resonance pole further supports the interpretation of
the $G(3900)$ as a $P$-wave $D\bar{D}^{*}$ molecular state. This molecular picture is also corroborated by Ref.~\cite{Chen:2025gxe}, where the Bethe--Salpeter equation is solved for
the $D\bar{D}^{*}$ interaction. Taken together, these studies provide indirect evidence in favor of interpreting the $G(3900)$ as a genuine resonant state.

Observing the $G(3900)$ in hadronic production processes, rather than through intermediate states or the $e^{+}e^{-}\to D\bar{D}$ reaction, would provide compelling evidence for its
interpretation as a genuine resonance. Kaon-induced reactions at facilities such as OKA@U-70~\cite{Obraztsov:2016lhp}, SPS@CERN~\cite{Velghe:2016jjw}, and the newly commissioned
AMBER@CERN~\cite{Quintans:2022utc}, together with future high-luminosity programs at colliders like the FCC~\cite{Tomas:2020trw}, provide promising opportunities for the discovery of
exotic hadrons.  Motivated by these prospects, we investigate the production cross section of $G(3900)$ in the reaction $K^- p \to D_s^- \Lambda_c^+ G(3900)^0$, including initial-state
interactions (ISIs) in the $K^- p$ system.  If the $G(3900)$ is a genuine resonance rather than a dynamical effect specific to $e^{+}e^{-}\to D\bar{D}$, it should also appear as a clear
enhancement in invariant-mass spectra in $K^{-}p$ collision. Its observation would therefore provide an important and independent test of its existence.

\begin{figure}[http]
\begin{center}
\includegraphics[bb=75 600 1150 710, clip, scale=0.55]{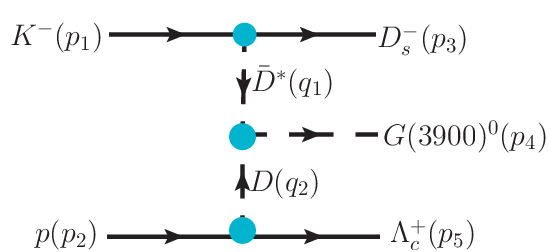}
\caption{Tree-level Feynman diagrams for the production of $G(3900)$ in the $K^- p \to D_s^- \Lambda_c^+ G(3900)^0$ reaction, where $p_1$, $p_2$, $p_3$, $p_4$, $p_5$, $q_1$, and $q_2$
denote the particle momenta. The arrows indicate the momentum flow.}\label{cc1}
\end{center}
\end{figure}

However, no dedicated search for the $G(3900)$ in $K^- p$ collisions through the reaction $K^- p \to D_s^- \Lambda_c^+ G(3900)^0$ has been reported to date.  To address this issue, we
propose to search for the $G(3900)$ in $K^- p $ collision via a central production mechanism with the corresponding Feynman diagram shown in Fig.~\ref{cc1}. Within this framework,
the process is formulated by treating the $G(3900)$ as a $P$-wave $D\bar{D}^{*}/D^{*}\bar{D}$ molecular state, with its production mechanism consequently dominated by $t$-channel exchange
of $D$ and $\bar{D}^{*}$ mesons.

This paper is organized as follows. In Sec.~\ref{Sec: formulism},  we will present the theoretical  formalism.  In Sec.~\ref{Sec: results}, the numerical result will be given, followed
by discussions and conclusions in the last section.

\section{FORMALISM AND INGREDIENTS}\label{Sec: formulism}
In this work, we investigate the production of the $G(3900)^0$ in the $K^- p \to D_s^- \Lambda_c^+ G(3900)^0$ reaction.  Treating the $G(3900)^0$ as a $P$-wave
$D\bar D^{*}$ molecular state implies a strong coupling to the $D\bar D^{*}$ channel, suggesting that it can be effectively produced via a $D\bar D^{*}G(3900)^0$ vertex. Therefore, we adopt
the central production mechanism, which has been widely applied to the production of mesons such as $\phi$, $f_0(1500)$, and $f_2(1275)$~\cite{Lebiedowicz:2011tp}. The corresponding Feynman
diagrams are shown in Fig.~\ref{cc1}, where the production is described via $t$-channel exchanges of $D$ and $\bar D^{*}$ mesons.  In this mechanism, high-energy $K^{-}p$ collisions first produce
a $D\bar D^{*}$ pair together with $\Lambda_c^{+}$ baryon and $D_s^{-}$ meson, followed by a strong final-state interaction of the $D\bar D^{*}$ system at low relative energies, which leads to the formation
of the $G(3900)^0$. Since the $s$- and $u$-channel contributions require the production of two additional $c\bar c$ pairs in $K^{-}p$ collisions, they are doubly suppressed by the Okubo--Zweig--Iizuka
(OZI) rule and can thus be safely neglected.

To evaluate the amplitudes corresponding to Fig.~\ref{cc1}, we first construct the effective Lagrangian densities for the relevant interaction vertices. The interaction vertex involving
 $\Lambda_c p D$ is taken from Refs.~\cite{Huang:2018wgr,Dong:2010xv}.
\begin{align}
\mathcal{L}_{\Lambda_c N D} &= i g_{DN\Lambda_c}\,\bar{\Lambda}_c \gamma_5 N D + \text{H.c.}
\end{align}
where the coupling constant $g_{\Lambda_c N D} = -13.98$ is obtained from SU(4)-invariant Lagrangians~\cite{Dong:2010xv}, expressed in terms of $g_{\pi NN} = 13.45$ and $g_{\rho NN} = 6.0$.
Here, $N$, $D$, and $\Lambda_c$ represent the nucleon, $D$-meson, and $\Lambda_c^+$ baryon fields, respectively.  For the $K^- D^* D_s^-$ vertex, we employ the effective Lagrangian~\cite{Hofmann:2005sw}
\begin{equation}
\mathcal{L}_{PPV} = \frac{i}{4} g_h \langle[\partial_\mu P, P] V^\mu\rangle \label{eq2},
\end{equation}
where $\langle \cdots \rangle$ denotes the trace over SU(4) flavor space. $P$ and $V^\mu$ denote the SU(4) pseudoscalar and vector meson fields, respectively. Their explicit forms are given by
\begin{equation}
P = \sqrt{2}
\begin{pmatrix}
\frac{\pi^0}{\sqrt{2}} + \frac{\eta}{\sqrt{6}} + \frac{\eta'}{\sqrt{3}} & \pi^+ & K^+ & \bar{D}^0 \\
\pi^- & -\frac{\pi^0}{\sqrt{2}} + \frac{\eta}{\sqrt{6}} + \frac{\eta'}{\sqrt{3}} & K^0 & D^- \\
K^- & \bar{K}^0 & -\frac{2}{\sqrt{6}}\eta + \frac{\eta'}{\sqrt{3}} & D_s^- \\
D^0 & D^+ & D_s^+ & \eta_c
\end{pmatrix},
\end{equation}
and
\begin{equation}
V^\mu =
\begin{pmatrix}
\frac{1}{\sqrt{2}}(\rho^0 + \omega) & \rho^+ & K^{*+} & \bar{D}^{*0} \\
\rho^- & \frac{1}{\sqrt{2}}(-\rho^0 + \omega) & K^{*0} & D^{*-} \\
K^{*-} & \bar{K}^{*0} & \phi & D_s^{*-} \\
D^{*0} & D^{*+} & D_s^{*+} & J/\psi
\end{pmatrix}^\mu .
\end{equation}
Then we obtain
\begin{align}
\mathcal{L}_{D^* D_s K} = &\; \frac{i g_h}{2}\left( K^0 \partial_\mu D_s^- - D_s^- \partial_\mu K^0 \right) D^{*+\,\mu}\nonumber\\
                          &+ \frac{i g_h}{2}\left( K^+ \partial_\mu D_s^- - D_s^- \partial_\mu K^+ \right) D^{*0\,\mu} \nonumber \\
                          &- \frac{i g_h}{2}\left( \bar{K}^0 \partial_\mu D_s^+ - D_s^+ \partial_\mu \bar{K}^0 \right) D^{*-\,\mu}\nonumber\\
                          &- \frac{i g_h}{2}\left( K^- \partial_\mu D_s^+ - D_s^+ \partial_\mu K^- \right) \bar{D}^{*0\,\mu}.
\end{align}
The coupling $g_h$ is fixed from the strong decay width of $K^* \to K\pi$.  With the help of Eq.~\ref{eq2}, the two-body decay width $K^{*+} \to K^0 \pi^+$ is related to $g_h$ as
\begin{equation}
\Gamma(K^{*+} \to K^0 \pi^+) = \frac{g_h^2}{24\pi m_{K^{*+}}^2}\, P_{\pi K^*}^3 = \frac{2}{3}\,\Gamma_{K^{*+}},
\end{equation}
where $P_{\pi K^*}$ is the three-momentum of the pion in the rest frame of the $K^*$ meson. Using the experimental value of the total decay width $\Gamma_{K^{*+}} = 50.3 \pm 0.8~\mathrm{MeV}$
and the hadron masses~\cite{ParticleDataGroup:2024cfk}, we obtain $g_h = 9.11$.

In addition to the interaction vertices described above, we also introduce the effective Lagrangian for the $G(3900)D\bar{D}^{*}$ coupling~\cite{Zhang:2009gy},
\begin{equation}
\mathcal{L}_{G(3900) D \bar{D}^{*}}
=-i g_{G D \bar{D}^{*}}
\epsilon^{\alpha\beta\mu\nu}
\left(\partial_{\alpha}G_{\beta}\right)
\left(\partial_{\mu}\bar{D}^{*}_{\nu}\right)
D+\mathrm{H.c.}\, .
\end{equation}
This Lagrangian is employed by us to fit the experimental $e^{+}e^{-}\to D\bar{D}^{*}$ data~\cite{Belle:2006hvs}, including the contribution from the $G(3900)$ resonance, yielding the coupling constant
$g_{GD\bar{D}^{*}}=0.334~\mathrm{GeV}$~\cite{Huang:2025xvv}.

In evaluating the scattering amplitudes of the $K^- p \to D_s^- \Lambda_c^+ G(3900)^0$ reaction, form factors are introduced to account for the finite size of hadrons.
For the exchanged $D$ and $D^{*}$ mesons, we employ the commonly used monopole form factor
\begin{equation}
{\cal{F}}_i(q_i^2)=\frac{\Lambda_i^2-m_i^2}{\Lambda_i^2-q_i^2},
\qquad i=D,D^{*},
\end{equation}
where $q_i$ and $m_i$ denote the four-momentum and mass of the exchanged meson, respectively. The cutoff parameter $\Lambda_i$ is related to the hadron size and is parameterized as
$\Lambda_i=m_i+\alpha\Lambda_{\rm QCD}$,  with $\Lambda_{\rm QCD}=220~\mathrm{MeV}$.  The dimensionless parameter $\alpha$ encodes nonperturbative QCD effects at low energies and
cannot be calculated from first principles. It is therefore treated as a free parameter constrained by experimental data and will be discussed in detail below.

The propagator of the exchanged pseudoscalar $D$ meson is given by
\begin{equation}
G_D(q)=\frac{i}{q^2-m_D^2},
\end{equation}
while that of the exchanged vector $D^{*}$ meson reads
\begin{equation}
G_{D^{*}}^{\mu\nu}(q)=\frac{i\left(-g^{\mu\nu}+q^\mu q^\nu/m_{D^{*}}^2\right)}{q^2-m_{D^{*}}^2},
\end{equation}
where $\mu$ and $\nu$ are the polarization indices of the vector meson.

Using the effective Lagrangians, propagators, and form factors introduced above, the scattering amplitudes corresponding to the Feynman diagrams shown in
Fig.~\ref{cc1} can be written as
\begin{align}
\mathcal{M}^{Born}&=\frac{-g_{DN\Lambda_c}g_{GD\bar{D}^{*}}g_h}{2}\bar{u}(p_5,s_5)\gamma^5u(p_2,s_2)\frac{1}{q_2^2-m_D^2}\nonumber\\
                  &\times{}\epsilon_{\alpha\beta\mu\nu}p_4^{\alpha}\epsilon^{*\beta}_{p_4,s_4}q_1^{\mu}\frac{-g^{\nu\eta}+q_1^{\nu}q_1^{\eta}/m^2_{\bar{D}^{*}}}{q^2_1-m^2_{\bar{D}^{*}}}(p_{3}+p_1)_{\eta}{\cal{F}},
       \label{eq:dkdkstar_rho}
\end{align}
where ${\cal{F}} = {\cal{F}}_{\bar{D}^{*}} {\cal{F}}_{D}$ represents the product of the form factors associated with the exchanged mesons.  $\bar{u}(p_5, s_{5})$
and $u(p_2, s_2)$ are the Dirac spinors of the outgoing $\Lambda_c^+$ and the initial proton, respectively, where  $s_{5}$ and $s_2$ their spin indices. $\epsilon^{*\beta}(p_4, s_4)$
denotes the polarization vector of the produced $G(3900)$, where $p_4$ and $s_4$ represent its four-momentum and spin, respectively.

\begin{figure}[http]
\begin{minipage}{\columnwidth}
\centering
\includegraphics[width=\linewidth]{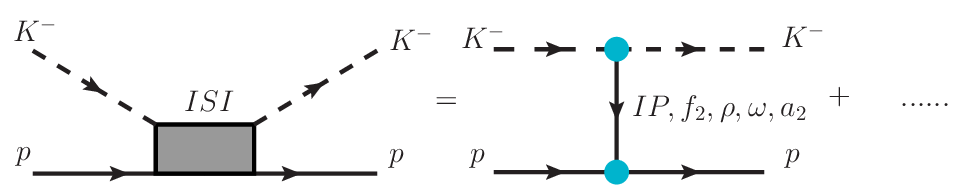}
\caption{The Feynman diagram illustrates the initial-state interaction (ISI) mechanism of the $K^-p \to K^-p$ reaction (the part on the left-hand side of the equality sign).
The right-hand side of the equality sign represents the tree-level scattering process, which includes the elastic and inelastic $K^-p$ interactions (indicated by the ellipses).}\label{ac2}
\end{minipage}
\end{figure}

To improve the reliability of our predictions, we include corrections to the Born amplitude in Eq.~\ref{eq:dkdkstar_rho}, arising from initial-state interactions (ISI) in the
$K^- p \to K^- p$ channel. These effects modify the scattering amplitude through both elastic and inelastic contributions, as illustrated in Fig.~\ref{ac2}, where the black square
denotes the full $K^- p \to K^- p$ amplitude including the tree-level elastic term and higher-order inelastic contributions represented by the ellipses.
We find that a simplified approach proposed in Ref.~\cite{Lebiedowicz:2011tp}, in which only the $K^- p \to K^- p$ elastic amplitude is considered and modeled via Pomeron and
Reggeon exchanges (see the right panel of Fig.~\ref{ac2}), is sufficient to reproduce the experimental data for $K^- p \to K^- p$ elastic scattering with high accuracy.
Therefore, we adopt this approach to account for the initial-state interactions in the $K^- p$ channel.

Within this framework, the total $K^- N \to K^- N$ amplitude is written as a sum of contributions from $\mathbb{P}$ (the Pomeron) and the $f_2$, $a_2$, $\omega$, and $\rho$
Reggeon exchanges~\cite{Lebiedowicz:2011tp}:
\begin{align}
T_{K^- N \to K^- N}(s,t) &= A_{\mathbb{P}}(s,t) + A_{f_2}(s,t) \pm A_{a_2}(s,t)\nonumber\\
                         &+ A_{\omega}(s,t) \pm A_{\rho}(s,t),
\end{align}
where $s$ is the squared center-of-mass energy and $t$ is the squared four-momentum transfer between the incoming and outgoing $K^-$ mesons. The upper (lower) signs correspond
to $K^- p \to K^- p$ ($K^- n \to K^- n$), respectively.  At sufficiently high center-of-mass energies, each contribution to the $K^- N \to K^- N$ amplitude can be parameterized as
\begin{equation}
A_i(s,t)= \eta_i\,s C^{\bar{K}N}_i\left(\frac{s}{s_0}\right)^{\alpha_i(t)-1}\exp\!\left(\frac{B^{\bar{K}N}_i}{2} t\right),
\end{equation}
where $i$ denotes the Pomeron ($\mathbb{P}$) and the Reggeons $f_2$, $a_2$, $\omega$, and $\rho$. The energy scale is taken as $s_0 = 1~\mathrm{GeV}^2$. The coupling constants $C^{\bar{K}N}_i$,
Regge trajectories $\alpha_i(t) = \alpha_i(0) + \alpha'_i t$, signature factors $\eta_i$, and slope parameters $B^{\bar{K}N}_i$ are taken from Ref.~\cite{Lebiedowicz:2011tp}, with numerical
values summarized in Table.~\ref{tab2}.
\begin{table}[htbp]
\centering
\caption{The parameters of the Pomeron and Reggeon exchanges were determined based on elastic and total cross section data in Ref.~\cite{Lebiedowicz:2011tp}.}
\label{tab2}
\scalebox{1.05}{
\begin{tabular}{@{\extracolsep{\fill}}lcccc@{}}
\hline
$i$ & $\eta_i$ & $\alpha_i(t)$ & $C_i^{\bar{K}N}$ (mb) & $B_i^{\bar{K}N}$ (GeV$^{-2}$) \\
\hline
$\mathrm{IP}$ & $i$ & $1.081 + (0.25~\mathrm{GeV}^{-2})\, t$ & 11.82 & 5.5 \\
$f_2$  & $-0.861+i$ & $0.548 + (0.93~\mathrm{GeV}^{-2})\, t$ & 15.67 & 4.0 \\
$\rho$  & $-1.162-i$ & $0.548 + (0.93~\mathrm{GeV}^{-2})\, t$ & 2.05 & 4.0 \\
$\omega$ & $-1.162-i$ & $0.548 + (0.93~\mathrm{GeV}^{-2})\, t$ & 7.055 & 4.0 \\
$a_2$  & $-0.861+i$ & $0.548 + (0.93~\mathrm{GeV}^{-2})\, t$ & 1.585 & 4.0 \\
\hline
\end{tabular}
}
\end{table}

After incorporating the initial-state interaction, the full amplitude can be written as~\cite{Lebiedowicz:2011tp}
\begin{equation}
\mathcal{M}_{\text{full}} = \mathcal{M}_{\text{Born}}+ \frac{i}{8\pi^2 s} \int d^2 \vec{k}_t \,T(s,k_t^2)\,\mathcal{M}_{\text{Born}}(s,k_t^2),
\end{equation}
where $\vec{k}_t$ denotes the transverse momentum transfer in the $K^- p \to K^- p$ reaction.  Using the above amplitude, we calculate the differential
cross sections for the processes $K^- p \to D_s^- \Lambda_c^+ G(3900)$  in the center-of-mass frame. Since these are $2 \to 3$ reactions, the cross
sections can be written in terms of the total amplitude as
\begin{align}
d\sigma &=\frac{m_p}{2\sqrt{(p_1 \cdot p_2)^2 - m_1^2 m_2^2}}\sum_{s_i,s_f} |\mathcal{M}_{\text{full}}|^2\frac{d^3 \vec{p}_3}{2E_3 (2\pi)^3} \frac{d^3 \vec{p}_4}{2E_4 (2\pi)^3}\nonumber\\
&\times\frac{m_{\Lambda_c^{+}}d^3 \vec{p}_5}{E_5 (2\pi)^3}(2\pi)^4 \delta^{(4)}(p_1+p_2-p_3-p_4-p_5),
\end{align}
where $E_3$, $E_4$, and $E_5$ are the energies of the $D_s^-$, $G(3900)$, and $\Lambda_c^+$, respectively.  $m_p$ and $m_{\Lambda_c^+}$ denote the masses of the initial proton and the
final-state $\Lambda_c^+$ baryon, respectively. $m_1$ and $m_2$ are the masses of the two incoming particles, with $m_2$ corresponding to the proton mass.  The sum $\sum_{s_i,s_f} |\mathcal{M}_{\text{full}}|^2$
denotes the sum over the spins of the initial- and final-state particles, where $s_i$ corresponds to the initial state and $s_f$ to the final state.  We adopt the following spin-sum
rules for the spin-$1/2$ baryons ($p$, $\Lambda_c^+$) and the spin-1 $G(3900)$ state, respectively,
\begin{align}
&\sum_{s_{2}} u(p_2,s_2)\bar{u}(p_2,s_2)= \frac{p\!\!\!/_2 + m_2}{2m_2}, \\
&\sum_{s_4} \epsilon^{\mu}(p_4,s_4)\epsilon^{*\nu}(p_4,s_4)= -g^{\mu\nu} + \frac{p_4^{\mu}p_4^{\nu}}{p_4^2}.
\end{align}

\section{RESULTS AND DISCUSSIONS}\label{Sec: results}
\begin{figure}[http]
\begin{center}
\includegraphics[bb=10 10 550 330, clip, scale=0.40]{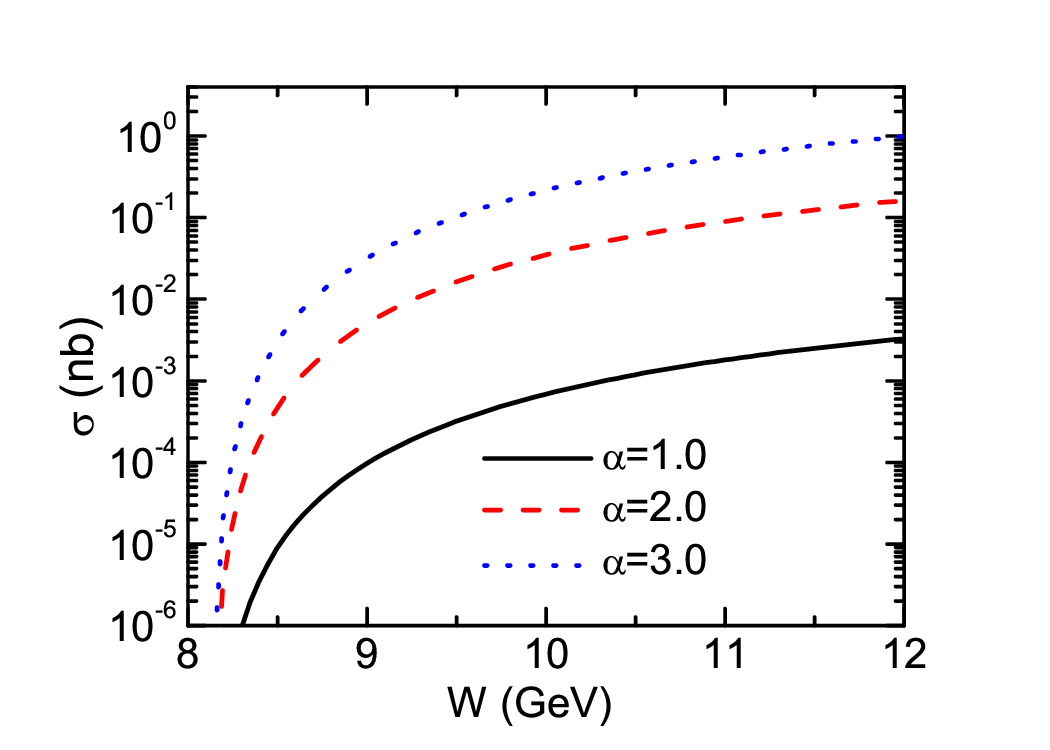}
\caption{The total cross section for the $K^- p \to D_s^- \Lambda_c^+ G(3900)^0$ reaction with different $\alpha$.
$W$ is the center-of-mass energy.}\label{cc2}
\end{center}
\end{figure}

Within our theoretical framework, the only parameter that significantly affects the predicted results is the cutoff parameter $\alpha$ associated with the form factor. As shown in Fig.~\ref{cc2},
in the absence of initial-state $K^- p$ interactions, we present the energy dependence of the cross section for the  reaction $K^- p \to D_s^- \Lambda_c^+ G(3900)^0$ at $\alpha = 1, 2,$ and $3$,
respectively, as a function of the center-of-mass energy $w$.  It is worth noting that the predicted cross sections are highly sensitive to the cutoff parameter $\alpha$.  To quantify this dependence,
we evaluate the cross section at the center-of-mass energy $W = 12~\text{GeV}$ as a representative case and investigate its variation in the range $\alpha = 1.0$--$3.0$.  The resulting cross sections
for the process $K^- p \to D_s^- \Lambda_c^+ G(3900)^0$ are found to increase from $0.00327~~\text{nb}$ at $\alpha=1.0$ to $0.163~~\text{nb}$ at $\alpha=2.0$, and further rise to $0.990~~\text{nb}$ at
$\alpha=3.0$, corresponding to an enhancement by a factor of approximately 303 between the minimum and maximum values.  In addition, the Fig.~\ref{cc2} also tell us that the increase in the cross section
as $\alpha$ varies from 1 to 2 is significantly larger than that obtained when $\alpha$ is increased from 2 to 3.  This pronounced sensitivity of the cross section to the cutoff parameter $\alpha$
indicates that a more tightly constrained determination of $\alpha$ would be highly desirable.

Although the value of $\alpha$ cannot be determined from first principles, it can be constrained using experimental data. In particular, the cutoff parameter $\alpha$ associated with the $D$ and $D^{*}$
meson exchange form factors has been fitted to experimental measurements~\cite{Belle:2007qxm,BaBar:2006qlj}. The detailed fitting procedure can be found in Ref.~\cite{Guo:2016iej}, where it was shown that
both $\alpha=1.5$ and $1.7$ provide a reasonable description of the data within uncertainties. Therefore, in the following calculations, we adopt these two values.
\begin{figure}[http]
\begin{center}
\includegraphics[bb=10 10 550 330, clip, scale=0.40]{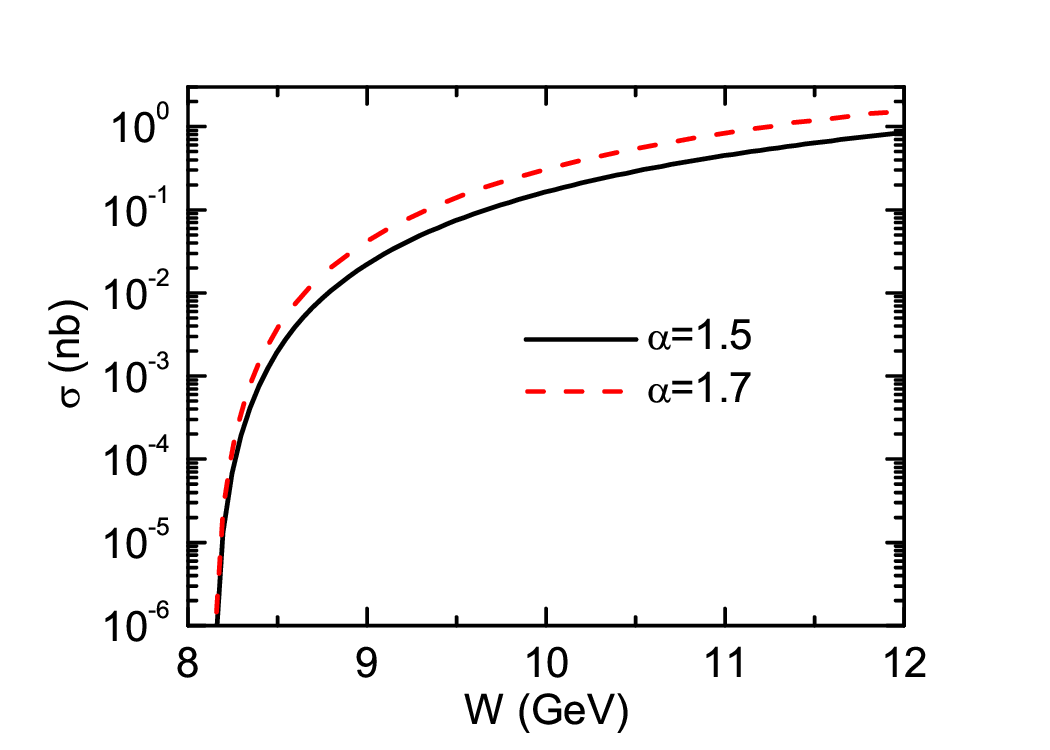}
\caption{Total cross sections of the $K^- p \to D_s^- \Lambda_c^+ G(3900)^0$ reaction with ISI included, plotted as functions of the center-of-mass energy. The black solid and red dashed curves correspond
to $\alpha=1.5$ and $1.7$, respectively.}\label{cc3}
\end{center}
\end{figure}

Using the obtained $\alpha$ values, the total cross sections for the $K^- p \to D_s^- \Lambda_c^+ G(3900)^0$ reaction, with the $K^-p$ initial-state interaction (ISI) taken into account, are presented
in Fig.~\ref{cc3} as functions of the center-of-mass energy $W$ from the reaction threshold to 12.0 GeV.  As shown in Fig.~\ref{cc3}, the total cross section exhibits a steep rise near the reaction threshold,
which can be attributed to the opening of the reaction phase space. As the center-of-mass energy increases further, the cross section continues to grow, but at a much slower rate than in the threshold region.

The results also indicate that the total cross section for $\alpha = 1.7$ is approximately twice as large as that for $\alpha = 1.5$. Moreover, the difference between the cross sections obtained with
$\alpha = 1.7$ and $\alpha = 1.5$ increases with the center-of-mass energy. For instance, at $W = 11$ GeV, the cross section is 0.83 nb for $\alpha = 1.7$, while it is 0.45 nb for $\alpha = 1.5$,
corresponding to a ratio of about 1.84. At $W = 11.5$ GeV, the corresponding values are 1.18 nb and 0.63 nb, respectively, giving a slightly larger ratio of approximately 1.87.

\begin{figure}[http]
\begin{center}
\includegraphics[bb=10 10 550 290, clip, scale=0.40]{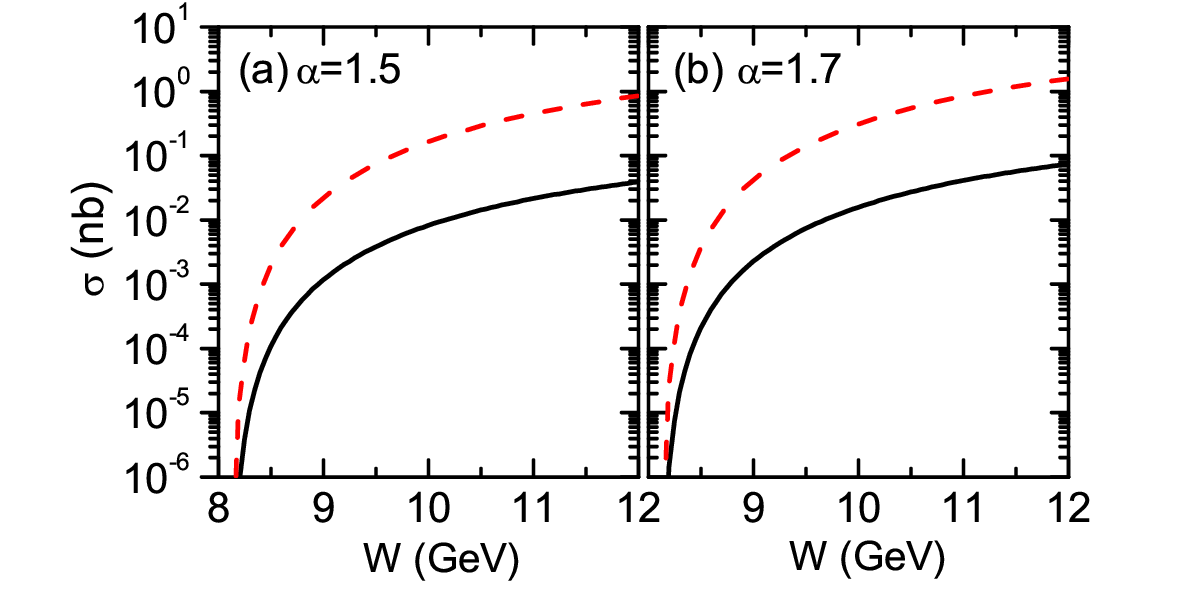}
\caption{Total cross sections for the $K^- p \to D_s^- \Lambda_c^+ G(3900)^0$ reaction as functions of the center-of-mass energy $W$, with and without the inclusion of initial-state interactions (ISI):
(a) the $\alpha=1.5$ case and (b) the $\alpha=1.7$ case.}\label{cc4}
\end{center}
\end{figure}

To assess the impact of the $\bar{K}N$ initial-state interaction (ISI), we compare the cross sections for the $K^- p \to D_s^- \Lambda_c^+ G(3900)^0$ reaction calculated with and without the inclusion of ISI,
as shown in Fig.~\ref{cc4}. Panels (a) and (b) correspond to the cases of $\alpha=1.5$ and $\alpha=1.7$, respectively. The black solid curves represent the Born-level predictions, whereas the red dashed curves
denote the full calculations including the $\bar{K}N$ ISI.

Evidently, the $\bar{K}N$ ISI plays an important role and leads to a substantial enhancement of the predicted cross sections over the entire energy
region considered.  To clarify this effect, we present a quantitative comparison of the cross sections at a fixed center-of-mass energy $W=11~\text{GeV}$. For $\alpha=1.5$, the cross section of the $K^- p \to D_s^- \Lambda_c^+ G(3900)^0$ reaction without including the $K^- p$ initial-state interaction is found to be only $0.021~\text{nb}$. In contrast, when the initial-state interaction is taken into account, the corresponding
cross section increases significantly to $0.45~\text{nb}$. This corresponds to an enhancement by approximately a factor of 21 due to the inclusion of the $K^- p$ ISI.  For $\alpha=1.7$, at a center-of-mass
energy $W=11~\text{GeV}$, the corresponding cross sections show a similar behavior, where the inclusion of the $K^- p$ initial-state interaction results in an enhancement by a factor of about 20, which is
slightly smaller than that obtained in the $\alpha=1.5$ case.
This enhancement may be understood as arising from re-scattering effects in the $\bar{K}N$ channel, which effectively modify the incoming wave function. This, in turn, increases the overlap between the $K^- p$
wave function and the production operator, thereby enhancing the effective transition amplitude and ultimately leading to a significant increase in the cross section.

\begin{figure}[http]
\begin{center}
\includegraphics[bb=10 10 550 330, clip, scale=0.38]{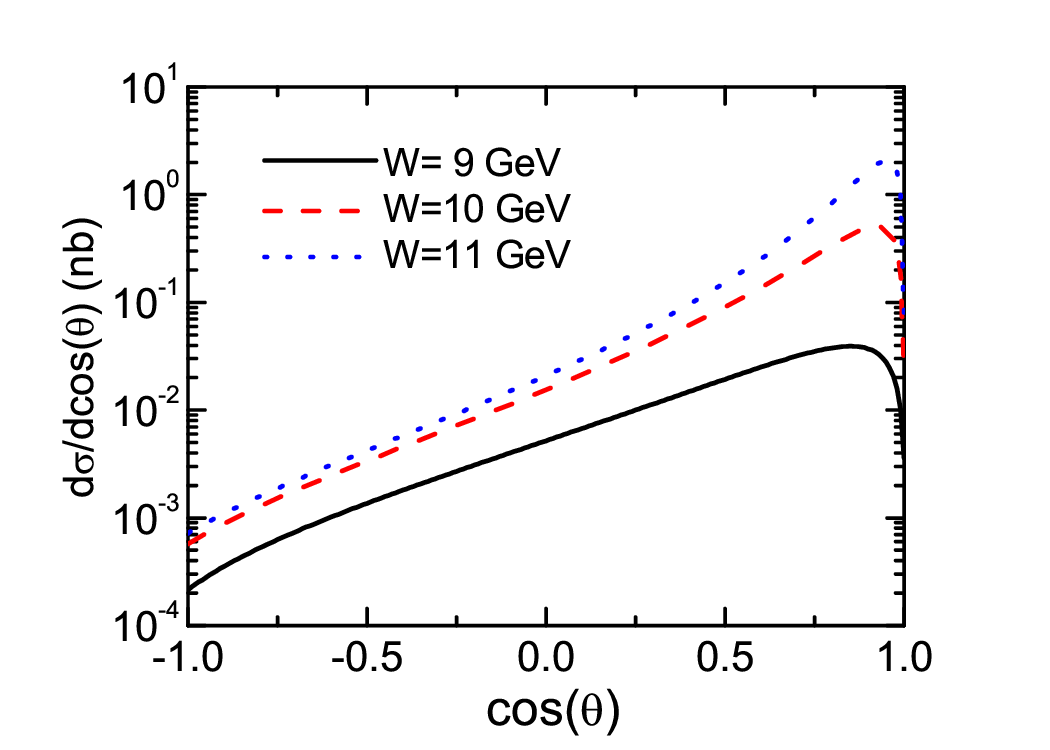}
\caption{ The $K^- p \to D_s^- \Lambda_c^+ G(3900)^0$ differential cross sections at different energies with $\alpha=1.5$. The black solid, red dashed, and blue dash-dotted curves correspond to
center-of-mass energies of $W = 9.0$, $10.0$, and $11.0$ ~$\mathrm{GeV}$, respectively.}\label{cc5}
\end{center}
\end{figure}

In addition to the total cross section, we also calculate the differential cross section for the $K^- p \to D_s^- \Lambda_c^+ G(3900)^0$ reaction as a function of the scattering angle of the outgoing
$G(3900)\Lambda_c^+$ system with respect to the beam direction in the center-of-mass frame. The numerical results are presented in Fig.~\ref{cc5} at $\alpha = 1.5$ for three representative center-of-mass
energies, $W = 9.0$, $10.0$, and $11.0~\mathrm{GeV}$. The differential cross section is found to increase with $W$ and exhibits a strong forward enhancement, decreasing rapidly with increasing scattering
angle. This behavior originates from the dominance of the $t$-channel exchange mechanism, where the reaction is mediated solely by $D$- and $D^{*}$-meson exchanges in the present model.

A closer inspection reveals that the differential cross section does not increase monotonically toward the extreme forward direction. Instead, at $W=9.0~\mathrm{GeV}$, it reaches a maximum around
$\cos\theta\simeq0.85$ before gradually decreasing. As the center-of-mass energy increases to $10.0$ and $11.0~\mathrm{GeV}$, the turnover point shifts progressively toward the forward direction,
occurring at approximately $\cos\theta\simeq0.92$ and $0.95$, respectively. This behavior indicates that the angular distribution becomes increasingly forward-peaked with increasing collision energy.
This structure is not caused by the initial-state interaction (ISI) but rather originates from the production mechanism employed in the present work. As illustrated in Fig.~\ref{cc1}, the presence of
two $t$-channel exchange contributions leads to a nonlinear dependence of the momentum transfer on the scattering angle. Consequently, the effective propagator structure deviates from the simple linear
form $1/(t_0 + A\cos\theta)$ and acquires higher-order angular dependences.

More explicitly, in the Born amplitude given in Eq.~(\ref{eq:dkdkstar_rho}), both $D^{*}$- and $D$-meson exchanges contribute simultaneously in the $t$-channel. Consequently, the differential cross
section is proportional to the product of two propagator factors,
\begin{equation}
\frac{1}{(t_{D^{*}} + A_{D^{*}}\cos\theta)} \cdot \frac{1}{(t_{D} + A_{D}\cos\theta)},
\end{equation}
which introduces nonlinear angular dependences beyond the simple $1/(A + B\cos\theta)$ behavior. This modifies the forward-peaking structure and leads to deviations from a strictly monotonic angular distribution.
Here,$t_{D^{*}} = m_1^2 + m_3^2 - 2E_1E_3$,$A_{D^{*}} = 2|\vec{p}_1||\vec{p}_3|$, where $E_1$ ($E_3$) and $|\vec{p}_1|$ ($|\vec{p}_3|$) denote the energies and magnitudes of the three-momenta of particles 1 (3),
respectively. The corresponding quantities $t_D$ and $A_D$ are obtained from $t_{D^{*}}$ and $A_{D^{*}}$ by the replacements $1 \rightarrow 2$ and $3 \rightarrow 5$.

\section{Summary}\label{sec:summary}
At present, there is still significant debate regarding whether the structure $G(3900)$ observed in the 2024 LHCb analysis should be interpreted as a genuine resonance. The main difficulty arises from the
fact that, in the process $e^{+}e^{-} \to D\bar{D}$, interference among conventional vector charmonia and threshold effects near the $D\bar{D}^{*}$ opening can also produce a structure compatible with the
observed signal.  Therefore, it is essential to investigate alternative production channels to further clarify the nature of $G(3900)$. In particular, studying different reactions where such interference
mechanisms are absent or significantly suppressed can help disentangle genuine resonant behavior from kinematic or dynamical effects, and thus provide a more reliable determination of whether $G(3900)$
corresponds to a true hadronic state.

In this work, we study the production of the $G(3900)$ resonance in the $K^- p \to D_s^- \Lambda_c^+ G(3900)^0$ reaction within an effective Lagrangian approach. The production mechanism is assumed to
proceed via a central production process, where the dominant contribution arises from $t$-channel $D^0$ and $D^{*0}$ exchanges. This choice is motivated by the interpretation of $G(3900)$ as a $\bar{D}^{*}D$
molecular state.  The coupling constant of $G(3900)$ to the $\bar{D}^{*}D$ channel is extracted from our previous fit to experimental data for the $e^{+}e^{-}\to \bar{D}^{*}D$ reaction~\cite{Huang:2025xvv}.
The $K^- p$ initial-state interaction (ISI) is incorporated through Pomeron and Reggeon exchanges~\cite{Lebiedowicz:2011tp}, which is found to enhance the production cross section by approximately one order
of magnitude, reaching a level of 0.1 nb.    In addition, we calculate the differential cross section for the $K^- p \to D_s^- \Lambda_c^+ G(3900)^0$ reaction. A strong forward enhancement is observed;
however, the interference of the two $t$-channel propagators ($\bar{D}^{*}$ and $D$ exchange) leads to a mild suppression in the extreme forward region ($\cos\theta \to 1$), resulting in a slight departure
from monotonic angular behavior.  These results indicate that the $G(3900)$ signal in this reaction could be accessible in future experiments. Its observation would be crucial for clarifying the nature of
$G(3900)$ and determining whether it corresponds to a genuine resonance state.

\section*{Acknowledgments}
This work was supported by the Sailing Plan Project of Yibin University (No. 2021QH06) and the National Natural Science Foundation of China under Grant No. 12005177. Yin Huang also acknowledges support from the Fundamental Research Funds for the Central Universities under Grant No. 2682026TPY011.

%\end{CJK*}
%

\begin{thebibliography}{23}%
%\cite{BESIII:2024ths}
\bibitem{BESIII:2024ths}
M.~Ablikim \textit{et al.} [BESIII],
%``Precise Measurement of Born Cross Sections for e+e-{\textrightarrow}DD{\textasciimacron} at s=3.80-4.95{\,}{\,}GeV,''
Phys. Rev. Lett. \textbf{133} (2024), 081901.


%\cite{Wang:2025dur}
\bibitem{Wang:2025dur}
X.~Wang, X.~Liu and Y.~Gao,
%``Colloquium: Hadron production in open-charm meson pairs at e+e- colliders,''
Rev. Mod. Phys. \textbf{98} (2026), 021001.



%\cite{BaBar:2006qlj}
\bibitem{BaBar:2006qlj}
B.~Aubert \textit{et al.} [BaBar],
%``Study of the Exclusive Initial-State Radiation Production of the D anti-D System,''
Phys. Rev. D \textbf{76} (2007), 111105.


%\cite{Belle:2007qxm}
\bibitem{Belle:2007qxm}
G.~Pakhlova \textit{et al.} [Belle],
%``Measurement of the near-threshold e+ e- ---{\ensuremath{>}} D anti-D cross section using initial-state radiation,''
Phys. Rev. D \textbf{77} (2008), 011103.


%\cite{Eichten:1979ms}
\bibitem{Eichten:1979ms}
E.~Eichten, K.~Gottfried, T.~Kinoshita, K.~D.~Lane and T.~M.~Yan,
%``Charmonium: Comparison with Experiment,''
Phys. Rev. D \textbf{21} (1980), 203.


%\cite{Zhang:2009gy}
\bibitem{Zhang:2009gy}
Y.~J.~Zhang and Q.~Zhao,
%``The Lineshape of psi(3770) and low-lying vector charmonium resonance parameters in e+ e- ---{\ensuremath{>}} D anti-D,''
Phys. Rev. D \textbf{81} (2010), 034011.


%\cite{Du:2016qcr}
\bibitem{Du:2016qcr}
M.~L.~Du, U.~G.~Mei{\ss}ner and Q.~Wang,
%``$P$-wave coupled channel effects in electron-positron annihilation,''
Phys. Rev. D \textbf{94} (2016), 096006.


%\cite{Lin:2024qcq}
\bibitem{Lin:2024qcq}
Z.~Y.~Lin, J.~Z.~Wang, J.~B.~Cheng, L.~Meng and S.~L.~Zhu,
%``Identification of the $G(3900)$ as the P-wave $D\bar{D}^*/\bar{D}D^*$ resonance,''
Phys. Rev. Lett. \textbf{133} (2024), 241903.


%\cite{Ye:2025ywy}
\bibitem{Ye:2025ywy}
Q.~Ye, Z.~Zhang, M.~L.~Du, U.~G.~Mei{\ss}ner, P.~Y.~Niu and Q.~Wang,
%``Resonance parameters of the vector charmoniumlike state G(3900),''
Phys. Rev. D \textbf{112} (2025), 016015.


%\cite{Nakamura:2023obk}
\bibitem{Nakamura:2023obk}
S.~X.~Nakamura, X.~H.~Li, H.~P.~Peng, Z.~T.~Sun and X.~R.~Zhou,
%``Global coupled-channel analysis of e+e-{\textrightarrow}cc{\textasciimacron} processes in s=3.75{\,}to{\,}4.7{\,}{\,}GeV,''
Phys. Rev. D \textbf{112} (2025), 054027.



%\cite{Husken:2024hmi}
\bibitem{Husken:2024hmi}
N.~H{\"u}sken, R.~F.~Lebed, R.~E.~Mitchell, E.~S.~Swanson, Y.~Q.~Wang and C.~Z.~Yuan,
%``Poles and poltergeists in e+e-{\textrightarrow}DD{\textasciimacron} data,''
Phys. Rev. D \textbf{109} (2024), 114010.


%\cite{Salnikov:2024wah}
\bibitem{Salnikov:2024wah}
S.~G.~Salnikov and A.~I.~Milstein,
%``Production of D(*)D{\textasciimacron}(*) near the thresholds in e+e- annihilation,''
Phys. Rev. D \textbf{109} (2024), 114015.



%\cite{Cao:2025okk}
\bibitem{Cao:2025okk}
J.~Cao, W.~C.~Zhang, Z.~L.~She, A.~K.~Lei, J.~P.~Zhang, H.~Zheng, D.~M.~Zhou, Y.~L.~Yan, Z.~Q.~Wang and B.~H.~Sa,
%``Charmonium-like exotic hadron productions in e+e- collisions at the BESIII energy with the PACIAE model,''
Phys. Rev. D \textbf{112} (2025), 014033.


%\cite{Huang:2025rvj}
\bibitem{Huang:2025rvj}
Y.~Huang and X.~Chen,
%``Direct Evidence for the $\bar{D}D^*/D\bar{D}^*$ Molecular Nature of $G(3900)$ Through Triangular Singularity Mechanisms,''
[arXiv:2501.10992 [hep-ph]].
%5 citations counted in INSPIRE as of 12 Jun 2026


%\cite{Huang:2025xvv}
\bibitem{Huang:2025xvv}
Y.~Huang and X.~Chen,
%``Criterion for the Existence of the $G(3900)$ Resonance,''
[arXiv:2509.06353 [hep-ph]].


%\cite{Lu:2025zae}
\bibitem{Lu:2025zae}
J.~L.~Lu, M.~Song, P.~Wang, J.~Y.~Guo, G.~Li and X.~Luo,
%``The bound and resonant states of $D^{(*)}D^{(*)}$ and $D^{(*)}{\bar{D}}^{(*)}$ with the complex scaling method,''
Eur. Phys. J. C \textbf{85} (2025), 920.



%\cite{Chen:2025gxe}
\bibitem{Chen:2025gxe}
X.~X.~Chen, Z.~M.~Ding and J.~He,
%``Pole trajectories from S- and P-wave DD{\textasciimacron}* interactions,''
Phys. Rev. D \textbf{111} (2025), 11.



%\cite{Obraztsov:2016lhp}
\bibitem{Obraztsov:2016lhp}
V.~Obraztsov [OKA],
%``High statistics measurement of the $K^+ \to \pi^0 e^+\nu$(Ke3) decay formfactors,''
Nucl. Part. Phys. Proc. \textbf{273-275} (2016), 1330-1333.


%\cite{Velghe:2016jjw}
\bibitem{Velghe:2016jjw}
B.~Velghe [NA62-RK and NA48/2],
%``$K^¡À ¡ú ¦Ð^¡À¦Ã¦Ã$ Studies at NA48/2 and NA62-RK Experiments at CERN,''
Nucl. Part. Phys. Proc. \textbf{273-275} (2016), 2720-2722.



%\cite{Quintans:2022utc}
\bibitem{Quintans:2022utc}
C.~Quintans [AMBER],
%``The New AMBER Experiment at the CERN SPS,''
Few Body Syst. \textbf{63} (2022), 72.


%\cite{Tomas:2020trw}
\bibitem{Tomas:2020trw}
R.~Tom{\'a}s, J.~Keintzel and S.~Papadopoulou,
%``Emittance growth from luminosity burn-off in future hadron colliders,''
Phys. Rev. Accel. Beams \textbf{23} (2020), 031002.


%\cite{Lebiedowicz:2011tp}
\bibitem{Lebiedowicz:2011tp}
P.~Lebiedowicz and A.~Szczurek,
%``$pp \to pp K^{+}K^{-}$ reaction at high energies,''
Phys. Rev. D \textbf{85} (2012), 014026.



%\cite{Huang:2018wgr}
\bibitem{Huang:2018wgr}
Y.~Huang, C.~j.~Xiao, Q.~F.~L{\"u}, R.~Wang, J.~He and L.~Geng,
%``Strong and radiative decays of $D\Xi$ molecular state and newly observed $\Omega_c$ states,''
Phys. Rev. D \textbf{97} (2018), 094013.


%\cite{Dong:2010xv}
\bibitem{Dong:2010xv}
Y.~Dong, A.~Faessler, T.~Gutsche, S.~Kumano and V.~E.~Lyubovitskij,
%``Radiative decay of $\Lambda_c(2940)^+$ in a hadronic molecule picture,''
Phys. Rev. D \textbf{82} (2010), 034035.


%\cite{Hofmann:2005sw}
\bibitem{Hofmann:2005sw}
J.~Hofmann and M.~F.~M.~Lutz,
%``Coupled-channel study of crypto-exotic baryons with charm,''
Nucl. Phys. A \textbf{763} (2005), 90-139.


%\cite{ParticleDataGroup:2024cfk}
\bibitem{ParticleDataGroup:2024cfk}
S.~Navas \textit{et al.} [Particle Data Group],
%``Review of particle physics,''
Phys. Rev. D \textbf{110}, 030001 (2024).


%\cite{Belle:2006hvs}
\bibitem{Belle:2006hvs}
K.~Abe \textit{et al.} [Belle],
%``Measurement of the near-threshold e+ e- ---{\ensuremath{>}} D(*)+- D(*)-+ cross section using initial-state radiation,''
Phys. Rev. Lett. \textbf{98} (2007), 092001.


%\cite{Guo:2016iej}
\bibitem{Guo:2016iej}
X.~D.~Guo, D.~Y.~Chen, H.~W.~Ke, X.~Liu and X.~Q.~Li,
%``Study on the rare decays of $Y(4630)$ induced by final state interactions,''
Phys. Rev. D \textbf{93} (2016), 054009.



\end{thebibliography}
\end{document}